\documentclass[english,aps,manuscript,endfloats,superscriptaddress]{revtex4}
\usepackage{epsfig,amsmath}
\usepackage{graphics}
\usepackage{graphicx}
\begin{document}
\author{Maxim F. Gelin}
\affiliation{Department of Chemistry,
         Technische Universit\"at  M\"unchen,
         D-85747 Garching, Germany}

\author{Dassia Egorova}
\affiliation{Institute of Physical Chemistry, 
Christian-Albrechts-Universit\"at zu Kiel, 
D-24098 Kiel, Germany}

\author{Wolfgang Domcke}
\affiliation{Department of Chemistry,
         Technische Universit\"at  M\"unchen,
         D-85747 Garching, Germany}

\title{Exact quantum master equation for a molecular aggregate coupled to
a harmonic bath}

\begin{abstract}
We consider a molecular aggregate consisting of $N$ identical monomers.
Each monomer comprises two electronic levels and a single harmonic
mode. The monomers interact with each other via dipole-dipole forces.
The monomer vibrational modes are bilinearly coupled to a bath of
harmonic oscillators. This is a prototypical model for the description
of coherent exciton transport, from quantum dots to photosynthetic
antennae. We derive an exact quantum master equation for such systems.
Computationally, the master equation may be useful for the testing of various
approximations  employed in theories of quantum transport. 
Physically, it offers a plausible explanation of the 
origins of long-lived coherent optical responses of molecular 
aggregates in  dissipative environments.
\end{abstract}
\maketitle

\section{Introduction}

Quantum systems coupled to a {}``bath'' of harmonic oscillators
are paradigmatic models in nonequilibrium quantum statistical mechanics
\cite{Weiss,Spohn,Leggett,may01,Yan05,Plenio10a}. The primary goal of the
theory of quantum dissipative systems is the derivation of so-called
master equations, which govern the dynamics of the reduced density
matrix of a {}``system'' of interest, starting from the Liouville
- von Neumann equation for the combined system+bath ensemble. A well
know example is Redfield theory (or modifications thereof \cite{Chernyak98}),
which is valid in the limit of weak system-bath coupling \cite{Weiss,Spohn,Leggett,may01,
Yan05,Plenio10a,Chernyak98,Redfield65}.
A master equation can also be derived in the limit of the strong system-bath
coupling \cite{Ankerhold01,Coffey07}. More advanced treatments have
lead to hierarchies of master equations \cite{Tanimura06a}, which
are valid beyond weak system-bath coupling and the Markovian approximation.
With these methods, the time evolution of the reduced density matrix
can be computed to high accuracy by the solution of truncated infinite
chains of equations for auxiliary density matrices.

There exist very few system-bath models for which {}``exact'' closed-form
quantum master equations have been derived. A master equation is called
exact if the time evolution of all operators representing observables
of the system is identical with the time evolution described by the
Liouville - von Neumann equation for the system+bath ensemble. An
obvious example is the so-called phase-noise case, in which the system
Hamiltonian commutes with the system-bath coupling \cite{hanngi08}.
Nontrivial exact quantum master equations are known for the free particle,
the harmonic oscillator, as well as for two or $N$ identical oscillators
bilinearly coupled to the harmonic bath \cite{reibold85,zhang92,yu96,grabert97,zhang07,hu08,hu08a}. 
The system-bath Hamiltonians
that are quadratic in  Bose or Fermi operators also allow 
for an analytical description of the system dynamics (see, e.g., \cite{Gardiner04,Prosen10,kosov09}).

In the present communication, we present the derivation of an exact
quantum master equation for a molecular aggregate consisting of $N$
identical molecular monomers. Each monomer comprises two electronic
levels and a single harmonic vibrational mode, which is coupled to
a harmonic bath. This model is prototypical for the description of
the transport of excitons, from quantum dots to photosynthetic 
antennae and DNA \cite{may01,Tanimura10,Mahan,QT,Renger06a,Mukamel09a}.
 Apart from curiosity
(as R. J. Baxter put it \cite{Baxter}, {}``... the model is relevant
and it can be solved, so why not to do so and see what it tells us?'')
our  motivation is threefold. 
Conceptually, the derived  master equation is, to our knowledge, the first 
exact master equation for a system 
 with  electronic inter-monomer couplings, intra-monomer
electron-vibrational couplings as well as vibrational dissipation.
None of these couplings are assumed to be weak.
Computationally, the exact master equation may be useful for the testing of various
approximations (e.g., weak system-bath coupling, or weak electron-vibrational
coupling, or weak inter-monomer coupling) which are frequently employed
in theories of quantum transport \cite{may01,Mahan,QT,Renger06a,Mukamel09a,Eisfeld11,Galperin07,Rainer11}. 
Fundamentally, the master equation offers a plausible explanation of the 
origins of long-lived coherent optical responses of molecular 
aggregates in  dissipative environments
\cite{fle07a,Moran09,scholes09,Wasie94,Osuka03,Struve97,fle07,kauff09,engel10,scholes10a}.

We use units in which $\hbar=1$.

\section{Derivation of the master equation}

Consider the total Hamiltonian $H$, which consists of the system
(S) Hamiltonian, the bath (B) Hamiltonian, and their coupling,\begin{equation}
H=H_{S}+H_{B}+H_{SB}.\label{H1}\end{equation}
 The system comprises $N$ electronic two-level systems each of which
possesses a vibrational mode:\begin{equation}
H_{S}=H_{ex}(B_{a}^{\dagger},B_{a'},|X_{j}-X_{j'}|)+\sum_{k=1}^{N}\left(\frac{P_{k}^{2}}{2M}+\frac{M\Omega^{2}X_{k}^{2}}{2}+\xi B_{k}^{\dagger}B_{k}X_{k}\right).\label{HS}\end{equation}
Here, $B_{a}^{\dagger}$ and $B_{a'}$ are the exciton creation and
annihilation operators obeying the Pauli commutation rules $[B_{a},B_{a'}^{\dagger}]=\delta_{aa'}(1-2B_{a}^{\dagger}B_{a'})$.
$X_{k}$, $P_{k}=-id/dX_{k}$, $M$, and $\Omega$ denote the
positions, momenta, masses, and frequencies of the harmonic oscillators.
The parameter $\xi$ controls the strength of the intra-monomer exciton-vibrational
coupling. The excitonic Hamiltonian $H_{ex}$ may depend parametrically
on the distances $|X_{j}-X_{j'}|$. We need not specify
$H_{ex}$, but we require that it conserves the number of excitons
and thus commutes with the number operator $\hat{N}$: \begin{equation}
[H_{ex},\hat{N}]=0,\,\,\,\hat{N}\equiv\sum_{k=1}^{N}B_{k}^{\dagger}B_{k}.\label{Nc}\end{equation}
We allow for variable couplings between the identical monomers, since
the monomers are fixed (e.g., in protein matrices) and interact via
(usually, dipole-dipole) forces, which depend on the relative positions
and orientations of the monomers. 

The system interacts with a harmonic bath via bilinear system-bath
coupling \begin{equation}
H_{SB}+H_{B}=\sum_{i=1}^{N_{B}}\left(\frac{p_{i}^{2}}{2m_{i}}+\frac{m_{i}\omega_{i}^{2}}{2}\sum_{k=1}^{N}(x_{i}-\frac{c_{i}X_{k}}{m_{i}\omega_{i}^{2}})^{2}\right).\label{HSB}\end{equation}
Here $x_{i}$, $p_{i}=-id/dx_{i}$, $m_{i}$, and $\omega_{i}$
denote the positions, momenta, masses, and frequencies of the bath
oscillators, and $c_{i}$ are the system-bath coupling coefficients.
All $N$ monomers are assumed to be identical, and the influence of
the bath (\ref{HSB}) on the system dynamics is determined by a single
spectral density \begin{equation}
g(\omega)=\frac{\pi}{2}\sum_{i=1}^{N_{B}}\frac{c_{i}^{2}}{m_{i}\omega_{i}}\delta(\omega_{i}-\omega).\label{Spec}\end{equation}
Otherwise, the masses $M$, frequencies  $\Omega$, electronic
couplings $\xi$, and system bath-coupling coefficients $c_{i}$ 
acquire a $k$-dependence, and an exact master equation cannot be
derived. However, the master equation derived below remains true for
different oscillators provided $M_{k}\sim\Lambda_{k}\sim\xi_{k}\sim c_{i,k}$
for any $k$.

In the description of  energy-transport and relaxation phenomena in excitonic
systems, the vibrational bath is
usually assumed to couple electronic degrees of freedom directly,
inducing fluctuations of the site energies \cite{Renger06a,Mukamel09a}.  
In the present study, as well as, e.g.,  in Refs. \cite{may01,all08,Gelin09a},
each monomer is coupled to the bath through its vibrational reaction mode.  As has
been shown in \cite{Onuchic85,Chernyak96,Burghardt09a,Burghardt09b,Nemeth10},
the two methods are, in principle, equivalent: By introducing an appropriate canonical
transformation, one can switch from one description to another by
incorporating the system modes into the bath or by singling-out several
(high-frequency) modes from the bath and treating them explicitly.
The present choice of the system-bath coupling through the monomer vibrational
modes is motivated  by three major reasons. First, it allows us to derive
an exact master equation. Second, the explicit treatment of high frequency
modes strongly coupled to electronic two-level systems allows us to
assume that the remaining bath modes are coupled to the monomers rather
weakly. This may become important for the extension of the present
theory beyond the system-bath Hamiltonian (\ref{HSB}). Third, the 
explicit consideration of  high-frequency vibrational modes facilitates 
the study of vibrationally-coherent effects in the system responses and spectroscopic 
signals (see, e.g., \cite{may01,all08,Gelin09a}).

Let us introduce new variables specifying the system oscillators,
the center-of-mass coordinate $R$ and the internal coordinates $Q_{j}$
\begin{equation}
R=\frac{1}{N}\sum_{j=1}^{N}X_{j};\,\,\, Q_{j}=X_{j+1}-X_{j},\,\,\, j=1,2,...,N-1.\label{XQ}\end{equation}
The transformation (\ref{XQ}) is described by the $N\times N$ matrix
$S$:\begin{equation}
Q_{j}=\sum_{k=1}^{N}S_{jk}X_{k},\,\, R=\sum_{k=1}^{N}S_{Nk}X_{k}.\label{XQ1}\end{equation}
The transformation back to the original coordinates $X_{k}$ is given
by the inverse matrix:\begin{equation}
X_{j}=\sum_{k=1}^{N-1}(S^{-1})_{jk}Q_{k}+(S^{-1})_{jN}R.\label{XQ2}\end{equation}
We were unable to derive an explicit analytical expression for $S^{-1}$
for arbitrary $N$ (for any finite $N$ this can be done numerically).
For the following derivations, it is sufficient to realize that $S^{-1}$
obeys the conditions\begin{equation}
\sum_{j=1}^{N}(S^{-1})_{jk}\equiv0,\,\,\,(k=1,2,...,N-1);\,\,\,(S^{-1})_{jN}\equiv1,\,\,\,(j=1,2,...,N),\label{XQ3}\end{equation}
which are elementary consequences of Eq. (\ref{XQ2}). According to
Eq. (\ref{XQ}), the original momenta are connected to the new momenta
$P_{R}\equiv-id/dR$ and $P_{Q,j}\equiv-id/dQ_{j}$ as
follows:\begin{equation}
P_{j}=\frac{1}{N}P_{R}-P_{Q,j}+P_{Q,j-1}.\label{PQ}\end{equation}
The change of the system vibrational variables (\ref{XQ}) is borrowed
from the theory of Gaussian polymers (see, e.g., \cite{Zwanzig74}).
It is linear, non-singular ($\det S=1$), but not canonical. The transformation
(\ref{XQ}) is just a technical tool, and we can return to the original
canonical $X_{j}$ representation at the end of the derivation. The
choice (\ref{XQ}) of the internal coordinates is natural for a linear
array of monomers. Other choices may be preferable in different situations.
The transformation (\ref{XQ}) can also be applied to the derivation
of the master equation for $N$ identical oscillators bilinearly coupled
to a harmonic bath, generalizing thereby the approach of Ref. \cite{hu08a}. 

By insertion of Eqs. (\ref{XQ2}) and (\ref{PQ}) into Eqs. (\ref{HS})
and (\ref{HSB}), we obtain the system Hamiltonians in the new variables
\begin{equation}
H_{S}=H_{S}^{(R)}+H_{S}^{(Q)},\label{HSxq}\end{equation}
\begin{equation}
H_{S}^{(R)}=\left(\frac{P_{R}^{2}}{2MN}+\frac{MN\Omega^{2}R^{2}}{2}+\xi N\hat{N}R\right),\label{HSr}\end{equation}
\begin{equation}
H_{S}^{(Q)}=H_{ex}(B_{a}^{\dagger},B_{a'},Q_{j})+\sum_{k=1}^{N}\frac{(P_{Q,k-1}-P_{Q,k})^{2}}{2M}+\sum_{k,l=1}^{N-1}G_{kl}\frac{M\Omega^{2}Q_{k}Q_{l}}{2}+\sum_{k=1}^{N}\sum_{l=1}^{N-1}\xi B_{k}^{\dagger}B_{k}(S^{-1})_{kl}Q_{l}\label{HSq}\end{equation}
(by definition, $P_{Q,0}=P_{Q,N}\equiv0$). Here $\hat{N}$ is the
exciton number operator (\ref{Nc}) and \begin{equation}
G_{kl}=\sum_{j=1}^{N}(S^{-1})_{jk}(S^{-1})_{jl}.\label{G}\end{equation}
The bath and the system-bath coupling Hamiltonians also split into
contributions associated with the variables $R$ and $Q$ 
\begin{equation}
H_{SB}+H_{B}=H_{SB}^{(R)}+H_{SB}^{(Q)}\label{HSB1}\end{equation}
where \begin{equation}
H_{SB}^{(R)}=\sum_{i=1}^{N_{B}}\left(\frac{p_{i}^{2}}{2m_{i}}+N\frac{m_{i}\omega_{i}^{2}}{2}(x_{i}-\frac{c_{i}R}{m_{i}\omega_{i}^{2}})^{2}\right)\label{HSBp}\end{equation}
and \begin{equation}
H_{SB}^{(Q)}=\sum_{k,l=1}^{N-1}G_{kl}\frac{\Lambda_{B}Q_{k}Q_{l}}{2}.\label{HSBQ}\end{equation}
Here \begin{equation}
\Lambda_{B}=\sum_{i=1}^{N_{B}}\frac{c_{i}^{2}}{m_{i}\omega_{i}^{2}},\label{ren}\end{equation}
so that $H_{SB}^{(Q)}$ is the bath-induced renormalization of
the potential energy (the so-called Lamb-shift). 
Combining the Hamiltonians, we have \begin{equation}
H=H_{\Sigma}^{(R)}+H_{\Sigma}^{(Q)};\label{Hu}\end{equation}
 \begin{equation}
H_{\Sigma}^{(R)}\equiv H_{S}^{(R)}+H_{SB}^{(R)},\,\,\,\, 
H_{\Sigma}^{(Q)}\equiv H_{S}^{(Q)}+H_{SB}^{(Q)}\label{Hsig}\end{equation}
(the subscript $\Sigma$ indicates total system+bath Hamiltonians). 
Due to the requirement (\ref{Nc}) the Hamiltonians (\ref{Hu}) commute
\begin{equation}
[H_{\Sigma}^{(R)},H_{\Sigma}^{(Q)}]=0.\label{com}\end{equation}

Equations  (\ref{Hu})-(\ref{com}) summarize the first main result of the present paper. 
In words, they show that the parent Hamiltonian (\ref{H1})
can be transformed into 
the Hamiltonian (\ref{Hu}), which is the sum of two mutually commuting Hamiltonians,  
$H_{\Sigma}^{(R)}$ and $H_{\Sigma}^{(Q)}$.
$H_{\Sigma}^{(Q)}$ depends on the internal coordinates $Q_{j}$ and is 
independent of the bath variables.  
The bath enters $H_{\Sigma}^{(Q)}$ only through the bath-induced potential, $H_{SB}^{(Q)}$. 
$H_{\Sigma}^{(R)}$ depends on the  
center-of-mass coordinate  $R$, through which the system is bilinearly 
coupled to the bath. The discrete degrees of freedom enter 
$H_{\Sigma}^{(R)}$ exclusively via the number operator  $\hat{N}$.
After the expansion in the eigenfunctions of  $\hat{N}$, the Hamiltonian $H_{\Sigma}^{(R)}$  
becomes harmonic. 
It is this latter fact which allows us to exactly integrate the bath out and  
derive an exact master equation.

For doing that, let us introduce the total (system+bath) density matrix $\tilde{\rho}(t)$,
 which obeys the Liouville-von Neumann equation \begin{equation}
\partial_{t}\tilde{\rho}(t)=-i[H_{\Sigma}^{(R)}+H_{\Sigma}^{(Q)},\tilde{\rho}(t)].\label{LvN}\end{equation}
Since the Hamiltonians $H_{\Sigma}^{(R)}$ and $H_{\Sigma}^{(Q)}$ commute, we can, 
following Ref. \cite{hu08}, introduce a new density
matrix \begin{equation}
\partial_{t}\tilde{\rho}^{(R)}(t)=-i[H_{\Sigma}^{(R)},\tilde{\rho}^{(R)}(t)],\label{LvN1}\end{equation}
 which is connected to the original density matrix via the unitary
transformation \begin{equation}
\tilde{\rho}(t)\equiv\exp\{-iH_{\Sigma}^{(Q)}t\}\tilde{\rho}^{(R)}(t)\exp\{iH_{\Sigma}^{(Q)}t\}.\label{roro}\end{equation}
Since $H_{\Sigma}^{(R)}$ (as well as $H_{\Sigma}^{(Q)}$) commutes
with the number operator $\hat{N}$, it can be expanded in the eigenfunctions
of $\hat{N}$: \begin{equation}
H_{\Sigma}^{(R)}\equiv\sum_{n=0}^{N}\left|n\right\rangle \left\langle n\right|H_{\Sigma,n}^{(R)}.\label{Hn}\end{equation}
Here\begin{equation}
H_{\Sigma,n}^{(R)}\equiv\left\langle n\right|H_{\Sigma}^{(R)}\left|n\right\rangle ,\,\,\,\hat{N}\left|n\right\rangle =n\left|n\right\rangle .\label{Hn1}\end{equation}
Therefore, Eq. (\ref{LvN1}) can be rewritten in the form
\begin{equation}
\partial_{t}\tilde{\rho}_{mn}^{(R)}(t)=-iH_{\Sigma,m}^{(R)}\tilde{\rho}_{mn}^{(R)}(t)+i\tilde{\rho}_{mn}^{(R)}(t)H_{\Sigma,n}^{(R)},\label{LvN1nm}\end{equation}
\[
\tilde{\rho}^{(R)}(t)\equiv\sum_{n,m=0}^{N}\left|n\right\rangle \left\langle m\right|\tilde{\rho}_{mn}^{(R)}(t).\]
Suppose that the density matrix at $t=0$ commutes with the number operator 
$\hat{N}$. Hence it can be expanded in the eigenfunctions of $\hat{N}$ as follows:  
 \begin{equation}
\tilde{\rho}^{(R)}(0)=\sum_{n=0}^{N}\left|n\right\rangle 
\left\langle n\right|\tilde{\rho}^{(R)}_{nn}(0).\label{Ro01}\end{equation}
$\tilde{\rho}^{(R)}_{nn}(0)$ are not limited to pure states: they 
can represent a linear combination of pure states belonging to the
same $n$. The initial condition (\ref{Ro01}) is adequate for many
practical purposes. Usually, the system is in its ground electronic
state ($n=0$) or can be promoted to its first excited electronic
state ($n=1$) by a laser pulse. Preparation of higher-order states
($n>1$) is also possible via multiple and/or strong laser pulses.
The level of description based on Eq. (\ref{Ro01}) is appropriate
for describing sequential (strong-pulse) spectroscopic signals, given
the pulses are temporally well-separated \cite{DW11}. If Eq. (\ref{Ro01})
is not fulfilled, we have to consider Eq. (\ref{LvN1nm}) for $n\neq m$.
In this case, $\tilde{\rho}_{mn}^{(R)}(t)$ is propagated via the 
Hamiltonian $H_{\Sigma,n}^{(R)}$ in the bra and via the Hamiltonian 
 $H_{\Sigma,m}^{(R)} \ne H_{\Sigma,n}^{(R)}$ in the ket, 
and  an exact master equation is impossible to derive.

Given the initial condition (\ref{Ro01}), we restrict ourselves to the consideration 
of  Eq. (\ref{LvN1nm}) with $n=m$,
\begin{equation}
\partial_{t}\tilde{\rho}_{nn}^{(R)}(t)=-i[H_{\Sigma,n}^{(R)},\tilde{\rho}_{nn}^{(R)}(t)].
\label{LvN1n}\end{equation}
The Hamiltonian $H_{\Sigma,n}^{(R)}$ (\ref{Hn1}) is the
sum of the system Hamiltonian $H_{S}^{(R)}$ (\ref{HSr}) (in
which me must substitute $\hat{N}$ by its eigenavalue $n$) and the
system-bath Hamiltonian $H_{SB}^{(R)}$ (\ref{HSBp}). Let us
now define the renormalized system oscillator mass $\overline{M}=NM$,
the renormalized bath oscillator frequencies $\bar{\omega}_{i}^{2}=N\omega_{i}^{2}$,
and the renormalized system-bath coupling coefficients $\bar{c}_{i}=Nc_{i}$,
as well as introduce the shifted center-of-mass coordinate $\overline{R}=R+n\xi/(M\Omega^{2})$,
and the shifted bath coordinates $\bar{x}_{i}= x_{i}+n\xi c_{i}^{2}/(M\Omega^{2}m_{i}\omega_{i}^{2})$.
After all these transformations, the Hamiltonian $H_{\Sigma,n}^{(R)}$
assumes the form
\begin{equation}
H_{\Sigma,n}^{(R)}=H_{\Sigma}^{(\overline{R})}\equiv\frac{P_{R}^{2}}{2\overline{M}}+\frac{\overline{M}\Omega^{2}\overline{R}^{2}}{2}+\sum_{i=1}^{N_{B}}\left(\frac{p_{i}^{2}}{2m_{i}}+\frac{m_{i}\bar{\omega}_{i}^{2}}{2}(\bar{x}_{i}-\frac{\bar{c}_{i}\overline{R}}{m_{i}\bar{\omega}_{i}^{2}})^{2}\right).\label{Hh}\end{equation}
Apparently, $H_{\Sigma}^{(\overline{R})}$ is $n$-independent; it
describes a harmonic oscillator of the mass $\overline{M}$ and frequency
$\Omega$ bilinearly coupled to the harmonic bath. The Liouville -
von Neumann equation (\ref{LvN1n}) thus becomes \begin{equation}
\partial_{t}\tilde{\rho}^{(\overline{R})}(t)=-i[H_{\Sigma}^{(\overline{R})},\tilde{\rho}^{(\overline{R})}(t)],\label{Lh}\end{equation}
$\tilde{\rho}^{(\overline{R})}(t)$ being the total density matrix
in the transformed variables. According to Refs. \cite{reibold85,zhang92,yu96,grabert97},
the bath can be integrated out in Eq. (\ref{Lh}) exactly, yielding
the master equation for the reduced density matrix $\rho^{(\overline{R})}(t)\equiv\textrm{Tr}_{B}\{\tilde{\rho}^{(\overline{R})}(t)\}$:\begin{equation}
\partial_{t}\rho^{(\overline{R})}(t)=-i[H_{S}^{(\overline{R})},\rho^{(\overline{R})}(t)]+\overline{\Re}(t)\rho^{(\overline{R})}(t).\label{Redp}\end{equation}
Here $H_{S}^{(\overline{R})}\equiv P_{R}^{2}/(2\overline{M})+\overline{M}\Omega^{2}\overline{R}^{2}/2$
and the dissipation operator is defined as follows: 
\begin{eqnarray}
\overline{\Re}(t)\rho^{(\overline{R})}(t)\equiv-ia(t)[\overline{R}^{2},
\rho^{(\overline{R})}(t)]-ib(t)[\overline{R},\{P_{R},\rho^{(\overline{R})}(t)\}] \notag \\ 
+c(t)[\overline{R},[P_{R},\rho^{(\overline{R})}(t)]]-d(t)[\overline{R},[\overline{R},\rho^{(\overline{R})}(t)]].\label{Red}
\end{eqnarray}
In the above equation, $\{.,.\}$ denotes the anti-commutator, and
$a(t),\, b(t),\, c(t),\, d(t)$ are real functions which are explicitly
defined in \cite{reibold85,zhang92,yu96,grabert97}; they are determined
through the spectral density of the renormalized bath introduced in Eq. (\ref{Hh}):
\begin{equation}
\bar{g}(\omega)=\frac{\pi}{2}\sum_{i=1}^{N_{B}}\frac{\bar{c}_{i}^{2}}{m_{i}\bar{\omega}_{i}}
\delta(\bar{\omega}_{i}-\omega)=Ng(\frac{\omega}{\sqrt{N}})\label{Spec1}\end{equation}
($g(\omega)$ is given by Eq. (\ref{Spec})). Returning to the original
variables $R$ yields \begin{equation}
\partial_{t}\rho_{nn}^{(R)}(t)=-i[H_{S,n}^{(R)},\rho_{nn}^{(R)}(t)]+\Re_{n}(t)\rho_{nn}^{(R)}(t).\label{Redp1}\end{equation}
Here $\rho_{nn}^{(R)},$ $H_{S,n}^{(R)}$, and $\Re_{n}(t)$ are obtained
from $\rho^{(\overline{R})}$, $H_{S}^{(\overline{R})}$, and $\overline{\Re}(t)$
by the replacement of $\overline{R}$ with $R+n\xi/(M\Omega^{2})$.

Combining Eqs. (\ref{LvN1}) and (\ref{roro}) and using Eq. (\ref{Red}),
we obtain the desired exact master equation for the reduced (system)
density matrix $\rho(t)\equiv\textrm{Tr}_{B}\{\tilde{\rho}(t)\}$:
\begin{equation}
\partial_{t}\rho(t)=-i[H_{S}^{(R)}+H_{\Sigma}^{(Q)},\rho(t)]+\Re(t)\rho(t),\label{MEe}\end{equation}
 $\Re(t)\equiv\sum_{n}\left|n\right\rangle \Re_{n}(t)\left\langle n\right|$.
Equivalently, the master equation can be rewritten as 
\begin{equation}
\partial_{t}\rho(t)=-i[H_{S}+H_{SB}^{(Q)},\rho(t)]+\Re(t)\rho(t),\label{ME}\end{equation}
$H_{SB}^{(Q)}$ (Eq. (\ref{HSBQ})) being the bath-induced renormalization
of the system potential. Eqs. (\ref{MEe}) and (\ref{ME}) can be
transformed back to the original $X_{i}$-representation, if desired.
Very similar master equations can be derived assuming that the operators
$B_{a}^{\dagger}$ and $B_{a'}$ obey either the Bose or the Fermi
commutation relations. 

From the computational point of view, $H_{SB}^{(Q)}$ cannot be factorized
into commuting sub-Hamiltonians for the $Q_{k}$. For small $N$ (see
\cite{Engel08} for the vibronic trimer) a suitable matrix representation
of $H_{SB}^{(Q)}$ can be introduced. Alternatively, one can invoke
the generalized Fulton-Gouterman transformation \cite{Wagner84,Wagner96},
which diagonalizes $H_{SB}^{(Q)}$ in the electronic Hilbert space,
reducing the problem to the numerical evaluation of eigenvalues and
eigenfunctions of the transformed Hamiltonian.

\section{Discussion}

The master equation (\ref{ME}) is mathematically exact
and equivalent to the Liouville - von Neumann equation with the initial
Hamiltonian (\ref{H1})-(\ref{HSB}). 
According to the  master equation (\ref{ME}), the density matrix in
the $Q$ subspace, $\rho^{(Q)}(t)\equiv\textrm{Tr}_{(R)}\{\rho(t)\}$,
does not experience dissipation. Indeed, \begin{equation}
\partial_{t}\rho^{(Q)}(t)=-i[H_{\Sigma}^{(Q)},\rho^{(Q)}(t)]\label{ME1}\end{equation}
because the contributions due to $H_{S}^{(R)}$ and $\Re(t)$ are
traced out (compare with Ref. \cite{hu08a}). The decoherence-free
subspace \cite{Whaley98} hence spans the entire $Q$ subspace.
Physically, this is a striking manifestation of quantum interference.
Analogous effects have recently been studied in \cite{Plenio09}.

The presence of the decoherence-free $Q$ subspace cannot be 
ubiquitous in real aggregates. The key simplifying
assumption is that all monomers are coupled to a single harmonic bath
(the Hamiltonian (\ref{HSB})). More generally, we can assume that
each monomer is coupled to its own bath,\begin{equation}
H_{SB}+H_{B}=\sum_{i=1}^{N_{B}}\sum_{k=1}^{N}\left(\frac{p_{i,k}^{2}}{2m_{i}}+\frac{m_{i}\omega_{i}^{2}}{2}(x_{i,k}-\frac{c_{i}X_{k}}{m_{i}\omega_{i}^{2}})^{2}\right).\label{HSBc}\end{equation}
The influence of these baths on the system dynamics is determined
by the spectral density $g_{jk}(\omega)$ ($1\leq j,k\leq N$ enumerate
different monomers). Apparently, the single bath (\ref{HSB}) corresponds
to $g_{jk}(\omega)=g(\omega)$ given by Eq. (\ref{Spec}) and is thus equivalent to  
$N$ fully correlated baths. 
Uncorrelated baths yield $g_{jk}(\omega)=g(\omega)\delta_{jk}$.
Partially correlated baths can also be defined by introducing the
correlation parameter $0\leq\gamma\leq1$ as follows: 
$g_{jk}(\omega)=g(\omega)(\delta_{jk}+\gamma(1-\delta_{jk}))$.
The fully correlated and uncorrelated baths correspond to $\gamma=$
$1$ and $0$, respectively. If $\gamma\neq1$, no decoherence-free
subspaces exist and the entire system relaxes to equilibrium. 
The above analysis suggests that the relaxation in the $R$ subspace  
has a characteristic rate $\nu \sim g(\omega)$, while the 
relaxation in the $Q$ subspace has a rate $\nu_{\gamma} \sim \nu(1-\gamma)$.

It is impossible to derive an exact master equation in the case of  partially 
correlated baths. However, the master equation (\ref{ME})
may be augmented with phenomenological dissipative operators which describe relaxation 
in the $Q$ subspace. This can be done, for example, by switching to the eigenvalue
representation in the $Q$ subspace and introducing the corresponding
Redfield operator. Another option is to use the dissipation operator
$-\nu_{\gamma}(1-\rho_{eq}^{(Q)}\textrm{Tr}_{(Q)}\{...\})$ \cite{gelin03}.
Here $\rho_{eq}^{(Q)}$ is the equilibrium Boltzmann distribution
corresponding to the Hamiltonian $H_{\Sigma}^{(Q)}$.

The problem of bath correlations can be analyzed from a microscopic 
perspective \cite{gelin09}. If the harmonic potentials in the Hamiltonians
$H_{SB}$ and $H_{B}$ are considered as linearizations of anharmonic
interaction potentials between the particles of the system and the
bath, one arrives at the Hamiltonian (\ref{HSB}), because the identical
particles of the system should interact via the same (in our case,
harmonic) potentials with the bath particles. The domain of validity
of such a primitive linearization of the interaction potentials is,
however, limited to short times. On the other hand, we can consider
the harmonic bath and the bilinear system-bath coupling in the spirit
of a normal mode analysis \cite{stratt1,stratt2}. Each particle
of the system then experiences different local potentials from the
bath particles, no matter whether the particles are identical or not.
This is tantamount to introducing different (possibly correlated)
local harmonic baths for each monomer, which ensure relaxation of
all ($Q$ and $R$) degrees of freedom. 
On physical grounds, one expects that intramolecular vibrational
baths should be close to the fully correlated limit, while 
environmental vibrational baths should be less correlated.

Uncorrelated baths are the default choice in many simulations, although
there exist strong experimental indications that this assumption is
not universally applicable \cite{fle07a,Moran09,scholes09,Tokmakoff02,Wolynes09},
and partially correlated baths are more appropriate for describing
coherent energy and exciton transport 
\cite{Renger02,Renger06,Cho05,Thorwart10,Ishizaki10,Plenio10,Kleinekath11,
Schulten11,Mukamel02,Abram10,Abram11,Tanimura07,Guzik10,Castro10,Whaley11}.
The present analysis suggests that relaxation of molecular aggregates 
in the $R$ and $Q$ subspaces is governed by  two different rates
$\nu$  and $\nu_{\gamma}$.  For fully correlated baths, $\nu_{\gamma}=0$.
One hence expects that $\nu_{\gamma}<\nu$ for partially correlated baths.
The  slow  relaxation in the $Q$ subspace  
manifests itself in slow decays of the populations and coherences
of the aggregate  density matrix. This  may be one of the 
reasons of the existence of long-lived coherent optical responses in dissipative excitonic systems 
\cite{fle07a,Moran09,scholes09,Wasie94,Osuka03,Struve97,fle07,kauff09,engel10,scholes10a}.

\section{Acknowledgments}

This work has been supported by the Deutsche Forschungsgemeinschaft
(DFG) through a research grant and the DFG-Cluster of Excellence {}``Munich-Centre
for Advanced Photonics'' (www.munich-photonics.de). We wish to thank
Sergy Grebenshchikov, Leah Z. Sharp, and Michael Thoss for
helpful discussions.

\end{document}